\documentclass[twocolumn]{aastex631}
\shorttitle{The gamma-ray emission of high redshift blazar B3 1343+451}
\shortauthors{Wu et al.}
\graphicspath{{./}{figures/}}
\begin{document}
\title{The Nature of the High-energy Gamma-Ray Radiation Associated with the High-redshift Blazar B3 1343+451}
\email{bzhdai@ynu.edu.cn}
\email{huwen.3000@jgsu.edu.cn}

\author[0000-0002-6292-057X]{Fan Wu}
\affiliation{School of Physics and Astronomy, Key Laboratory of Astroparticle Physics of Yunnan Province, Yunnan University, Kunming 650091, P. R. China\\}
\author[0000-0003-0429-8636]{Wen Hu}
\affiliation{Department of Physics, Jinggangshan University, Jian, 343009, P. R. China\\}
\author[0000-0001-7908-4996]{Benzhong Dai}
\affiliation{School of Physics and Astronomy, Key Laboratory of Astroparticle Physics of Yunnan Province, Yunnan University, Kunming 650091, P. R. China\\}

\begin{abstract}
High-redshift blazars are the most powerful extragalactic astrophysical sources ever detected in the high-energy $\gamma$-ray band. In this study, we present a temporal and spectral analysis of the high-redshift blazar B3 1343+451 based on 14 years of Fermi-LAT observations, spanning from 2008 August 4 to 2022 June 6 (MJD 54,686—59,733). We extract a seven day binned $\gamma$-ray light curve in the energy range 0.1--500 GeV and identify seven  outburst periods with a peak flux of $>4.32\times10^{-7} \rm ph \cdot cm^{-2} \cdot s^{-1}$. The highest seven day flux (above 100 MeV) reaches $(8.06\pm0.56)\times10^{-7} \rm erg \ cm^{-2} \ s^{-1}$ on MJD = 56,177.16, which is 10 times higher than the flux in the quiescent period. To understand the properties of distant blazar jets, we employ a standard one-zone leptonic scenario and model the multiwavelength spectral energy distributions of one quiescent and seven flaring periods. We find that the $\gamma$-ray spectrum is better reproduced if we assume that the dissipation region of the jet, $R_{\rm diss}$, is located within the molecular torus, where infrared emission is the dominant external photon field. We infer that the jets in higher-redshift blazars have larger power and kinetic energy, where the kinetic energy is significantly greater than the radiation power, and the jet production efficiency suggests that we need to lower the accretion efficiency. These results imply that B3 1343+451 may have a standard thin disk surrounding its massive black hole, and the jets of B3 1343+451 may not be fully explained by the Blandford--Payne  process.
\end{abstract}

\section{Introduction\label{sec1}}

Blazars are a subclass of active galactic nuclei (AGNs) characterized by the presence of a collimated relativistic jet closely aligned to the observer's line of sight, and the resulting Doppler boosting generates high apparent luminosities and rapid variability, especially at high energies. Based on the absence or presence of broad emission lines in their optical spectra, blazars are divided into two subclasses: BL Lacertae objects and flat spectrum radio quasars \citep[FSRQs;]{1991ApJ...374..431S}. BL Lacertae objects exhibit weak or undetected emission lines, while FSRQs display prominent emission lines.

The spectral energy distributions (SEDs) of blazars extend from radio to $\gamma$-rays with a typical double-peaked structure: a low-energy (LE) component peaking between infrared (IR) and X-rays, and a high-energy (HE) component with a peak at MeV--TeV $\gamma$-ray energies. The former is widely believed to be synchrotron radiation of highly relativistic electrons accelerated in the blazar jet \citep{1986ApJ...308...78L, 1989MNRAS.241P..43G}, whereas the latter is often thought to arise from inverse Compton (IC) scattering of the synchrotron photons produced within the jet by the same population of energetic electrons responsible for the synchrotron emission \citep[synchrotron self-Compton (SSC); e.g.,][]{1996ApJ...461..657B, 1985A&A...146..204G} and/or from up-scattering of external photons entering the jet, i.e., external Compton (EC). In BL Lacertae objects, the SSC process is widely accepted as the dominant mechanism responsible for the $\gamma$-ray emission. The EC mechanism likely plays a dominant role, particularly in those sources where the HE component largely dominates the overall SED; this is usually observed in FSRQs. However, there is no consensus on the source of the external photons involved in the EC process. Two widely discussed external photon fields are the thermal IR radiation from the parsec-scale molecular torus \citep[(MT); e.g.,][]{2000ApJ...545..107B} and the optical emission lines from the subparsec-scale broad-lined region \citep[(BLR); e.g.,][]{1994ApJ...421..153S,2009MNRAS.397..985G, 2018MNRAS.477.4749C}.

Currently, the fourth catalog of Fermi-LAT AGNs reports the detection of only 110 blazars with redshifts $z > 2$, and only 33 blazars with redshifts $z > 2.5$ \citep{2020ApJS..247...33A,2022ApJS..260...53A}. They typically exhibit a flat or rising X-ray spectrum in the logarithmic plot of $\nu F_\nu$ versus $\nu$, especially in the hard X-ray band, have soft $\gamma$-ray time-averaged spectra ($\Gamma_\gamma>2.2$), and a $\gamma$-ray luminosity exceeding $10^{48}\ \rm ergs \cdot s^{-1}$. These high-redshift blazars are thought to harbor extremely massive black holes (BHs), with masses often in excess of $10^9 M_\odot$, accompanied by an accretion disk with a luminosity exceeding $10^{46}\ \rm ergs \cdot s^{-1}$ \citep[e.g.,][]{2010MNRAS.405..387G,Paliya2020ApJ,2020MNRAS.498.2594S,Marcotulli2020ApJ}.

It has been noticed that the typical properties exhibited by the most extreme blazars can be simply comprehended in accordance with the \emph{blazar sequence} \citep{Fossati1998MNRAS,Prandini2022Galax}. Particularly, distant and powerful objects, on the one hand, are crucial for studying the physics of relativistic jets and the accretion--jet connection during the early Universe \citep{Ghisellini2013MNRAS,2011MNRAS.416..216V}. On the other hand, they can place useful constraints on the density of the extragalactic background light \citep[EBL; e.g.,][]{2008A&A...487..837F,Finke2010ApJ} and therefore improve our understanding of cosmological evolution \citep[e.g.,][]{Ackermann2012Sci,Fermi2018Sci,Desai2019ApJ,Finke2022ApJ}.

As a representative high-redshift source, B3 1343+451, located at redshift $z = 2.534$, is among the top 30 brightest blazars observed by Fermi-LAT, and has exhibited intense flaring activities in the GeV $\gamma$-ray band since 2008. Motivated by the high-quality $\gamma$-ray data accumulated during 2008--2022, we performed a detailed spectral and variability analysis of this source.

Specifically, $\gamma$-ray data combined with archival data in other bands have allowed us to place a strong constraint on the physical processes responsible for the $\gamma$-ray emission and the physical properties of the relativistic jets of a blazar that formed near the beginning of the Universe. Our primary goal is to understand its physical properties by means of theoretical SED modeling, with a major focus on the location of the HE emission region and the plasma.

The paper is organized as follows. The data reduction of the $\gamma$-ray data is presented in Section~\ref{sec2}. Section~\ref{sec3} describes our results. In Section~\ref{sec4.1}, we present our theoretical modelling of the broadband SED, the results of which are discussed in Section~\ref{sec4.2}. A discussion and conclusions are provided in Section~\ref{sec5}. Throughout this paper, we use a flat cosmology with $\rm H_0=70.5 \  km\cdot s^{-1} \cdot Mpc^{-1}$, $\Omega_{\rm m}=0.27$, and $\Omega_\Lambda=0.73$. For B3 1343+541, its luminosity distance is $\rm d_L\simeq21.3 \ Gpc$.

\begin{figure*}[!ht]
\centering
\includegraphics[width=1\textwidth]{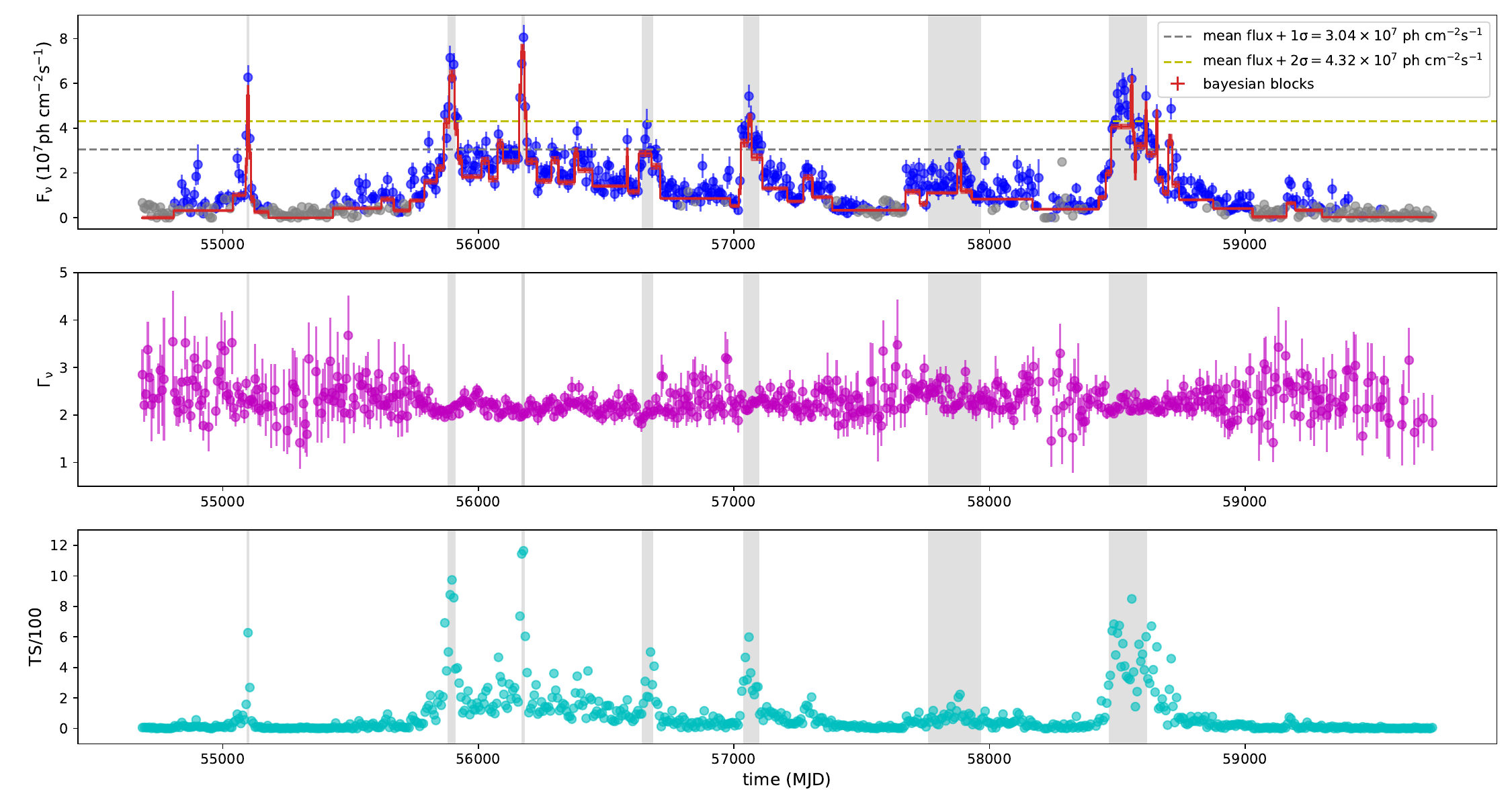}
\caption{The weekly binned $\gamma$-ray light curve and photon index of B3 1343+451 from 2008 August 4 (MJD 54,686) to 2022 June 6 (MJD 59,733). In the light curve, data points with $\rm TS<9$ are represented in gray and the two dashed lines represent flux thresholds. The bottom panel shows TS values for each bin. Shaded regions indicate the seven instances of active states studied in this work. The red lines represent the BB result with a FAP parameter $p_0=0.05$.
\label{fig1}}
\end{figure*}

\section{Fermi-LAT Observations and Data Reduction\label{sec2}}
Operating since 2008, Fermi-LAT has provided an exceptional view of the $\gamma$-ray sky, imaging the entire sky every three hours \citep{2009ApJ...697.1071A}. In the current paper, the Fermi-LAT data accumulated between 2008 August 04 and 2022 June 06 (MJD = 54,686—59,733) were downloaded and analyzed using the Fermi ScienceTools version {\tt 2.0.8} with the {\tt P8R3\_SOURCE\_V3} instrument response function, in a region of interest (ROI) centered on the $\gamma$-ray position of B3 1343+451 and with a radius of $15^\circ$. In the 100 MeV--500 GeV range, we selected events with a high probability of being photons ({\tt evclass = 128} and {\tt evtype = 3}) using the {\tt gtselect} tool.

A zenith angle cut of $90^\circ$ was applied to reduce contamination due to $\gamma$-rays from the Earth's limb. We also selected good time intervals in which the satellite was working in standard data-taking mode and the data quality was good. The current {\tt gtmktime} filter expression recommended by the LAT team in the Cicerone, \texttt{(DATA$\_$QUAL$>$0) \&\& (LAT$\_$CONFIG==1)}, was employed. For the $\gamma$-ray source in the whole ROI, a model file containing the spectral parameters was generated with the {\tt make4FGLxml.py} script based on the fourth Fermi-LAT source catalogue of $\gamma$-ray sources (4FGL); this produced an XML file containing all the sources with the user-defined ROI plus 10 degrees . In the likelihood fit, all the parameters of the sources lying outside the ROI were kept fixed to their values in 4FGL, while the normalization and the spectral parameters of the sources within the ROI were left to vary freely. Finally, the galactic and extragalactic diffuse $\gamma$-ray emission was parametrized using the {\tt gll\_iem\_v07.fits} and {\tt iso\_P8R3\_SOURCE\_V3\_v1.txt} models, respectively. Note that in this work, all data points have not been processed using the EBL absorption correlation; that is, we use the observed values after absorption.

\begin{figure*}[!ht]
\centering
\includegraphics[width=0.66\textwidth]{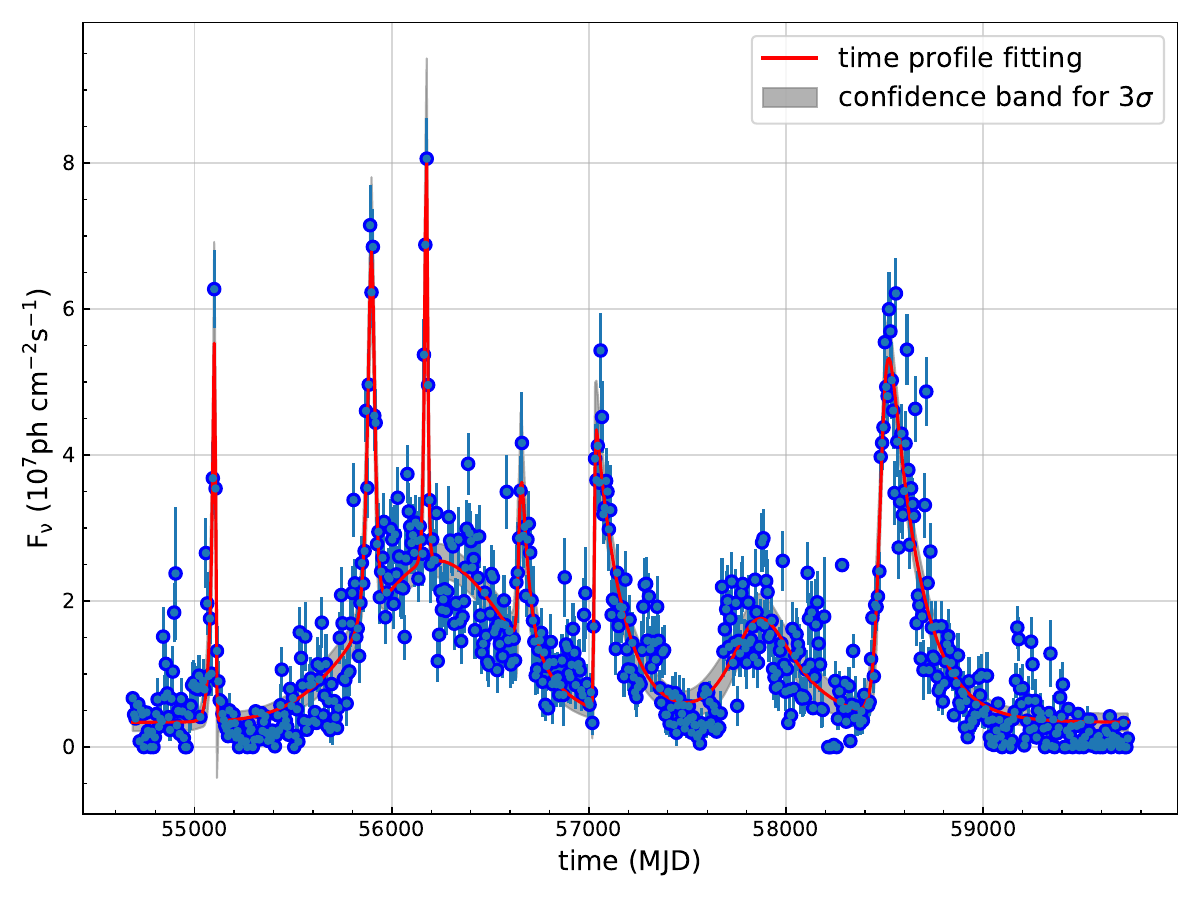}
\caption{Fitting the variability timescale obtained from the global light curve. \label{fig2}}
\end{figure*}

\begin{deluxetable*}{ccccccc}[!ht]
\tablenum{1}
\tablecaption{Best-fitting Results of the Weekly Binned $\gamma$-ray Light Curve of B3 1343+451 \label{tab1}}
\tablewidth{0pt}
\tablehead{
\colhead{Period} & \colhead{$F_{\rm i}$} & \colhead{$t_{\rm i}$} & \colhead{$T_{\rm r,i}$} & \colhead{$T_{\rm d,i}$}&\colhead{flare intervals}  \\
\cline{2-6}
 \colhead{ }     & \colhead{$\rm 10^{-7} \ ph \ cm^{-2} \ s^{-1}$} &\colhead{MJD}&\colhead{day} &\colhead{day}&\colhead{MJD}
}
\startdata
  Flare 1 & $4.94\pm 0.49$ & $55099.16$ & $4.47\pm 0.75$ & $5.05\pm 0.79$& $55094.69-55104.21$ \\
  Flare 2 & $4.63\pm 0.38$ & $55901.74$ & $21.68\pm3.94$ & $10.90\pm2.39$& $55880.06-55912.64$ \\
  Flare 3 & $5.71\pm 0.45$ & $56177.16$ & $9.04\pm1.01$ & $5.01\pm0.72$& $56168.12-56182.17$ \\
  Flare 4 & $2.37\pm 0.26$ & $56660.16$ & $19.99\pm3.84$ & $25.08\pm4.29$& $56640.17-56685.24$ \\
  Flare 5 & $3.08\pm 0.21$ & $57059.15$ & $23.30\pm2.91$ & $40.99\pm4.11$& $57035.85-57100.14$ \\
  Flare 6 & $0.79\pm 0.12$ & $57843.01$ & $83.47\pm25.18$ & $124.12\pm31.13$& $57759.54-57967.13$\\
  Flare 7 & $3.05\pm 0.17$ & $58485.68$ & $18.05\pm3.00$ & $132.07\pm11.20$ & $58467.63-58617.75$ \\
\enddata
\tablecomments{Columns 2--5 denote the peak flux, peak time, rise timescale, decay timescale, and associated errors of each flare obtained by fitting, respectively. Column 6 shows the flare intervals.}
\end{deluxetable*}

\section{Gamma-Ray Data Results\label{sec3}}

\subsection{Gamma-Ray Variability\label{sec3.1}}

One of the characteristics of blazars is a rapid change of flux with time. In Fig.~\ref{fig1}, we shown the $\gamma$-ray light curve, the photon index, and TS value obtained with a bin size of seven days. The photon index ($\Gamma_{\rm \nu}$)  is unsteady: when TS$>$9, the hardest and softest values are $1.42\pm0.41$ and $3.52\pm0.56$, respectively. The maximum one week averaged flux (above 100 MeV) was observed on MJD 56,177.16, reaching $(8.06\pm0.56)\times10^{-7} \rm erg \ cm^{-2} \ s^{-1}$, which is the highest weekly binned flux ever observed from this source. The photon index corresponding to this highest flux is $2.03\pm0.05$.

Identifying flares relying solely on naive visual inspection is often subject to bias, so we adopted the Bayesian Blocks (BB) algorithm to identify flares \citep{2013ApJ...764..167S} with a false alarm probability (FAP) parameter of $p_0=0.05$. BB can assist in identifying the location and extent of significant peaks in the light curve. If the average flux of the central block is significantly higher than the average fluxes of both the preceding and succeeding blocks \citep[a flux rise of at least a factor of two;]{2016A&A...593A..91A}, and the flux values in the central block are greater than the flux thresholds, it suggests the presence of a flare in this region.

The gray and yellow dashed lines in Fig.~\ref{fig1} represent the flux thresholds, which are the mean of the global light curve plus one or two times the standard deviation \citep{2020A&A...642A.129S}. Six peaks have flux points exceeding the threshold marked by the yellow dashed line, while one peak has a flux point exceeding the threshold of the gray dashed line, bringing the total number of suspected flares to seven.

To determine further the flare intervals quantitatively, we conducted time profile fitting using asymmetric flare templates \citep{2013ApJ...764...57D} for each peak, i.e.,\citep{2010ApJ...722..520A}:

\begin{equation}
f_i(t)=F_0+\frac{2F_{\rm i}}{\exp(\frac{t_{\rm i}-t}{T_{\rm r,i}})+\exp(\frac{t-t_{\rm i}}{T_{\rm d,i}})}
\end{equation}

where $F_0$ is the quiescent flux (can be a function or a constant), $F_{\rm i}$ is the flux at $t_{\rm i}$, representing the approximate flare amplitude, $t_{\rm i}$ is the time corresponding to the approximate flux peak (if it is a symmetric flare, $t_{\rm i}$ is the time corresponding to the actual flux peak), and $T_{\rm r,i}$ and $T_{\rm d,i}$ represent the rise and the decay timescales of the flare, respectively. The light curve was fitted with the nonlinear optimization python package {\tt lmfit}\footnote{\url{https://lmfit.github.io/lmfit-py/}}, and we defined the flare intervals as $(t_{\rm i}-T_{\rm r,i})$ $(t_{\rm i}+T_{\rm d,i})$ \citep{2021MNRAS.504..416S}, as shown in Fig.~\ref{fig2} and Table~\ref{tab1}.

\begin{figure*}[!ht]
\centering
\includegraphics[width=1\textwidth]{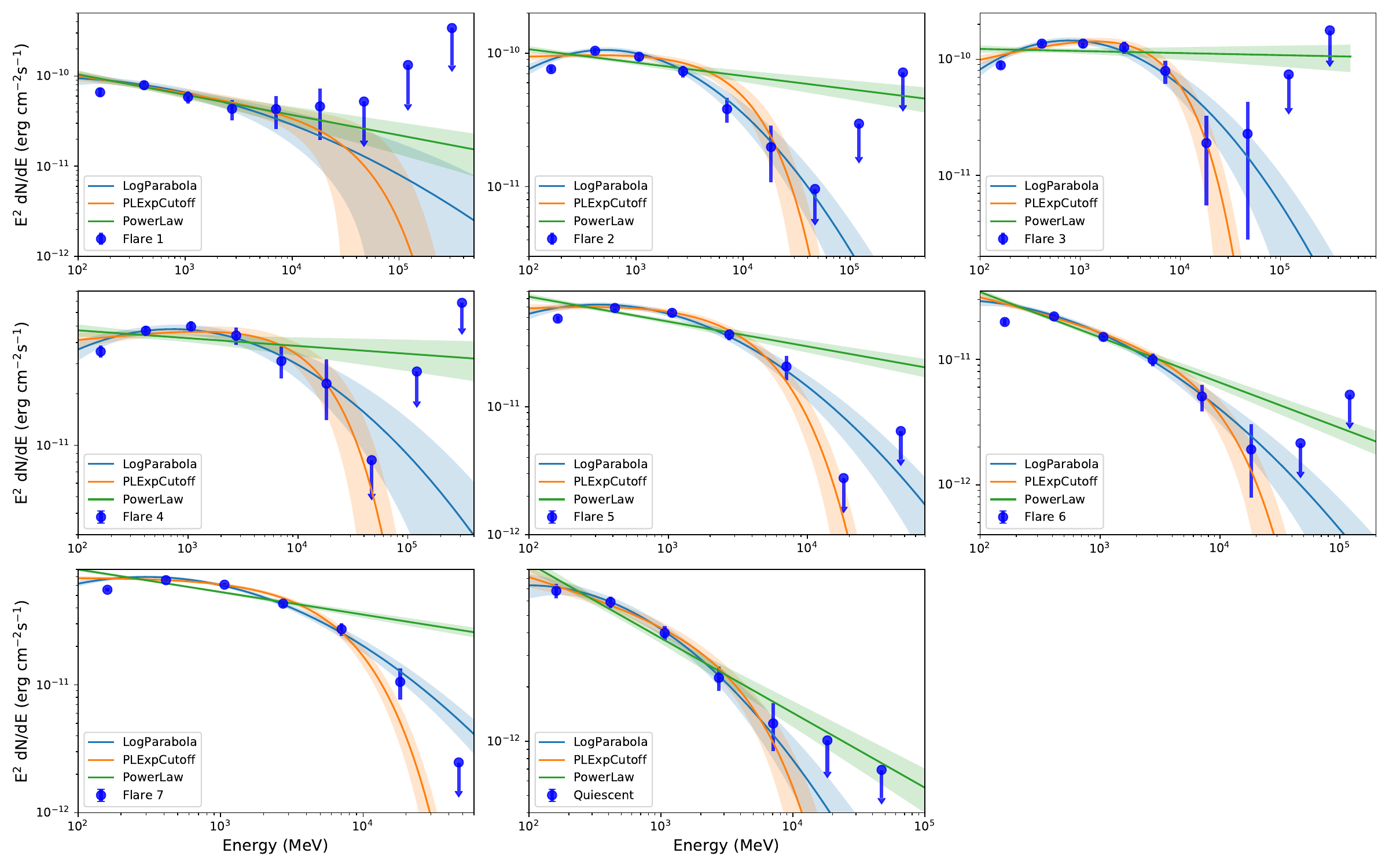}
\caption{$\gamma$-ray spectra of B3 1343+451 during the quiescent and seven active states. The data points are fitted with PL (green), PLC (orange), and LP (blue) models (see text). \label{fig3}}
\end{figure*}

\begin{deluxetable*}{ccccccccc}[!ht]
\tabletypesize{\scriptsize}
\tablewidth{0pt}
\tablenum{2}
\tablecaption{The Three Models Fit to the Seven Flare Periods and One Quiescent Period \label{tab2}}
\tablehead{\colhead{Model}&\colhead{Period}& \colhead{$\rm F_{0.1-500GeV}$ }  &  \colhead{ $N_0$  }    &\colhead{$ \rm \alpha$}&\colhead{$ \rm \beta$}&\colhead{$\rm E_{cut}$}& \colhead{TS}&\colhead{$\rm TS_{\rm curve}$}\\
\cline{3-9}
\colhead{}     & \colhead{}     &\colhead{$\rm 10^{-7} \ ph \ cm^{-2}s^{-1}$}&\colhead{$\rm 10^{-10} \ ph \ cm^{-2}s^{-1}MeV^{-1}$}& \colhead{ $\rm $}&\colhead{} & \colhead{  GeV  }       &\colhead{}&\colhead{}
}
\startdata
         & Flare 1 & $5.31\pm0.43$ &$2.00\pm0.15$& $2.22\pm0.07$ & $-$ &$-$ &615.94 &$-$ \\
         & Flare 2 & $6.07\pm0.24$ &$2.49\pm0.09$& $2.10\pm0.03$ & $-$ &$-$ &3151.07&$-$\\
         & Flare 3 & $7.52\pm0.38$ &$3.25\pm0.15$& $2.02\pm0.04$ & $-$ &$-$ &2078.09&$-$ \\
PL    & Flare 4 & $2.82\pm0.16$ &$1.20\pm0.06$& $2.05\pm0.04$ & $-$ &$-$ &1540.93&$-$ \\
         & Flare 5 & $3.79\pm0.15$ &$1.46\pm0.05$& $2.19\pm0.03$ & $-$ &$-$ &2749.66&$-$\\
         & Flare 6 & $1.58\pm0.06$ &$0.53\pm0.02$& $2.36\pm0.03$ & $-$ &$-$ &1972.86&$-$ \\
         & Flare 7 & $4.24\pm0.09$ &$1.65\pm0.03$& $2.18\pm0.02$ & $-$ &$-$ &9101.98&$-$ \\
         &quiescent& $0.43\pm0.03$ &$0.14\pm0.01$& $2.42\pm0.05$ & $-$ &$-$ &637.15 &$-$\\
\cline{1-9}
            & Flare 1 & $5.22\pm0.44$ &$2.11\pm0.20$& $2.19\pm0.08$ & $0.04\pm0.06$ & $-$ &617.09&-0.88\\
            & Flare 2 & $5.67\pm0.24$ &$2.88\pm0.13$& $1.99\pm0.04$ & $0.12\pm0.03$ & $-$ &3187.42&-29.70\\
            & Flare 3 & $6.98\pm0.38$ &$3.83\pm0.21$& $1.87\pm0.05$ & $0.14\pm0.03$ & $-$ &2112.78&-21.91\\
LP       & Flare 4 & $2.63\pm0.17$ &$1.29\pm0.07$& $1.94\pm0.06$ & $0.07\pm0.03$ & $-$ &1550.62&-7.41\\
            & Flare 5 & $3.55\pm0.15$ &$1.68\pm0.07$& $2.10\pm0.04$ & $0.12\pm0.03$ & $-$ &2778.55&-24.72\\
            & Flare 6 & $1.52\pm0.07$ &$0.58\pm0.03$& $2.32\pm0.04$ & $0.08\pm0.03$ & $-$ &1982.53&-8.85\\
            & Flare 7 & $4.02\pm0.10$ &$1.86\pm0.04$& $2.10\pm0.02$ & $0.10\pm0.02$ & $-$ &9173.91&-60.44\\
            &quiescent& $0.39\pm0.03$ &$0.15\pm0.01$& $2.33\pm0.06$ & $0.11\pm0.05$ & $-$ &642.15&-6.76\\
\cline{1-9}
         & Flare 1 & $5.26\pm0.43$ & $2.06\pm0.18$ & $2.19\pm0.09$ & $-$ &$40.6\pm27.5$&616.73&-0.59 \\
         & Flare 2 & $5.87\pm0.24$ & $2.74\pm0.13$ & $1.96\pm0.05$ & $-$ &$11.5\pm4.2$&3172.38&-17.33 \\
         & Flare 3 & $7.14\pm0.38$ & $3.82\pm0.24$ & $1.79\pm0.07$ & $-$ &$6.6\pm2.2$&2109.10&-20.23\\
PLC & Flare 4 & $2.70\pm0.16$ & $1.26\pm0.07$ & $1.93\pm0.06$ & $-$ &$19.1\pm9.6$&1550.79&-8.23\\
         & Flare 5 & $3.58\pm0.15$ & $1.80\pm0.11$ & $1.94\pm0.06$ & $-$ &$4.4\pm1.2$&2786.71 &-32.78\\
         & Flare 6 & $1.54\pm0.07$ & $0.59\pm0.03$ & $2.24\pm0.06$ & $-$ &$9.3\pm4.0$&1983.69&-10.09\\
         & Flare 7 & $4.06\pm0.09$ & $1.91\pm0.06$ & $1.99\pm0.03$ & $-$ &$6.9\pm1.2$&9185.38&-72.32 \\
         &quiescent& $0.40\pm0.03$ & $0.16\pm0.01$ & $2.22\pm0.09$ & $-$ &$5.8\pm2.9$&644.37&-8.51\\
\enddata
\tablecomments{Column 3 is the $\gamma$-ray flux. Column 4 is the spectral normalization factor when $\rm E_0=478.48 \ MeV$. Columns 5--7 are the fitting parameters for the three models. Columns 8--9 are the TS and $\rm TS_{curve}$ values obtained for the observing period, respectively.}
\end{deluxetable*}

\subsection{Gamma-Ray Spectral Analysis\label{sec3.2}}

Based on variability, the SEDs of both the quiescent \citep{2020Ap.....63..334S} (MJD=55,125—55,722) and flaring periods have been studied individually (see Fig.~\ref{fig3}). Three spectral models are adopted to analyze the $\gamma$-ray spectral shape. The first is a simple power law (PL):

\begin{equation}
dN/dE\propto \left(\frac{E}{E_0}\right)^{-\alpha},
\end{equation}

where $E_0$ is the scale energy and $\alpha$ is the PL index.

The second form is a PL with an exponential cutoff (PLC):

\begin{equation}
\frac{dN}{dE}\propto\left(\frac{E}{E_0}\right)^{-\alpha}\exp\left(-\frac{E}{E_{\rm c}}\right),
\end{equation}

where $E_{\rm c}$ is the cutoff energy.

The third form is a log-parabola (LP):

 \begin{equation}
 dN/dE\propto \left(\frac{E}{E_0}\right)^{-\alpha-\beta\log(E/E_0)},
 \end{equation}

where $\beta$ is the curvature index that describes the curvature around the peak. In our analysis, the value of $E_0$ is fixed at 478.48 MeV.

The best-fitting parameters and the spectral normalization factors are given in Table~\ref{tab2}; the integrated fluxes (above 0.1 GeV) derived using each spectral model are also reported. The values of the curvature index derived with the LP model range from $0.04\pm0.06$ to $0.12\pm0.03$, and the cutoff energy inferred with the PLC model for the different flux states is between $4.4\pm1.2$ and $40.6\pm27.5$ GeV.

We have examined the curvature in the spectrum by estimating $\rm TS_{curve}=2(\log \mathcal{L}_{\rm PLC/LP} - \log\mathcal{L}_{\rm PL})$ \citep{2012ApJS..199...31N}, where $\mathcal{L}_{\rm PLC/LP}$ and $\mathcal{L}_{\rm PL}$ represent the likelihood values estimated for the PLC (or LP) and PL models, respectively. More negative $\rm TS_{curve}$ values suggest better fits.

Our results demonstrate that the $\gamma$-ray spectra show a clear deviation from the PL model, and instead display curvature features. Both the PLC and LP models can better describe the $\gamma$-ray spectral shapes than the PL model, and the $\rm TS_{\rm curve}$ values do not differ much in these two cases. Considering that, we noted that flares 1, 2 and 3 are better fit with the LP, while flares 4, 5, 6, 7, and the quiescent period are represented best by the PLC. Similar results were also reported for other FSRQs by \citet{2017ApJ...844...62P} and \citet{2023MNRAS.520.2024K}.

\section{Broadband SEDs\label{sec4}}

We collected multiwavelength data through Space Science Data Center (SSDC) Sky Explorer\footnote{\url{http://tools.ssdc.asi.it/SED/}}. Broadband SEDs covering from IR wavelengths up to $\gamma$-ray energies were then constructed.

\subsection{Modelling the Broadband Emission\label{sec4.1}}

To interpret the broadband emission from B3 1343+451, we consider a standard one-zone leptonic model, including synchrotron emission, SSC, EC emission, and the $\gamma\gamma$ opacity due to interactions with the EBL field. The model is mostly designed to concentrate on emission from the optical/UV to $\gamma$-ray bands, rather than the radio flux, which may be mainly associated with the larger-scale jet structure.

For completeness, here we describe the model in brief. Henceforth, physical quantities with a prime are measured in the jet’s comoving frame, whereas the quantities without a prime are measured in the stationary AGN frame, unless specified otherwise. The emission region is assumed to be a spherical plasma blob of size $R'$, located at a distance of $R_{\rm diss}$ from a central BH of mass $M_{\rm BH}$, and filled with a population of relativistic electrons $N'_{\rm e}(\gamma')$ and a randomly oriented magnetic field of strength $B'$. The emission region moves relativistically with bulk Lorentz factor $\Gamma$ along the jet, oriented at a small angle $\theta_{\rm obs}\simeq\Gamma^{-1}$ with respect to our line of sight. The Doppler factor of the blob is approximated as $\delta_{\rm D}\sim\Gamma$.

Based on the analysis of the $\gamma$-ray spectral shape, we assume that the electron energy distribution (EED) follows an LP form:

\begin{equation}
\gamma'^2N_{\rm e}'(\gamma')=\gamma_{\rm pk}'^2 N_{\rm pk}' \left(\frac{\gamma'}{\gamma_{\rm pk}'}\right)^{-r \log (\gamma' / \gamma'_{\rm pk})}
\end{equation}

where $N_{\rm pk}'$ is the normalization constant at the peak Lorentz factor $\gamma_{\rm pk}'$ and $r$ is the spectral curvature parameter, which could be indicative of stochastic acceleration of electrons in the jet \citep[e.g.,][]{2006A&A...448..861M,2011ApJ...739...66T}.

The external radiation fields are comprised of optical/UV photons emitted directly from the accretion disk, with optical emission lines reprocessed by the BLR, and IR thermal radiation emitted by the MT.

The accretion disk is assumed to be a geometrically thin, optically thick Shakura--Sunyaev type and produces a multitemperature blackbody spectrum. It extends from $R_{\rm in,d}=3R_{\rm s}$ to $R_{\rm out,d}=10^3 R_{\rm s}$, where $R_{\rm s}\equiv2G M_{\rm BH} /c^2$ is the Schwarzschild radius and $G$ is the gravitational constant. Both the BLR and MT are assumed to be thin spherical shells located at distances $R_{\rm BLR}\simeq10^{17}\sqrt{L_{\rm disk}/10^{45}}$ cm and $R_{\rm MT}\simeq10^{18} \sqrt{L_{\rm disk}/10^{45}}$ cm, respectively, where $L_{\rm disk}$ is the total bolometric luminosity of the accretion disk. Their spectra are approximated as an isotropic blackbody peaking at $\sim10.2$ and $3.93k_{\rm B}T_{\rm MT}$ eV, respectively, where $k_{\rm B}$ is the Boltzmann constant and $T_{\rm MT}$ is the characteristic temperature of the MT. The energy of the BLR photons is based on the strong Ly$\alpha$ line in blazars, and we assume $T_{\rm MT}=10^3$ K for the MT radiation. Fractions of the accretion disk luminosity reprocessed by the BLR and the MT are adopted as $\tau_{\rm BLR}=0.1$ and $\tau_{\rm MT}=0.2$, respectively.

For B3 1343+451, we use the derived $L_{\rm disk}=4.8\times10^{45} \ \rm ergs\cdot s^{-1}$ and $M_{\rm BH}=10^9M_\odot$ from \citet{2020MNRAS.498.2594S}, where the accretion disk luminosity and the BH mass are estimated through modeling of the accretion disk spectrum over the optical/UV part of the SED during the quiescent period.

High-energy $\gamma$-ray photons traveling from the source towards the Earth will be absorbed by the pair-production process after interaction with the EBL. To account for the effect of EBL attenuation, the intrinsic $\gamma$-ray flux is converted to the observed flux by using the EBL model of \cite{Finke2010ApJ}. The observed flux is related to the intrinsic flux through:

\begin{equation}
f^{\rm obs}(\nu_{\rm obs})=f^{\rm int}(\nu)\exp\left[-\tau_{\rm EBL}(z,\nu_{\rm obs})\right]
\end{equation}

where $\nu=\nu_{\rm obs}(1+z)$.

\begin{figure*}[!ht]
\centering
\includegraphics[width=1\textwidth]{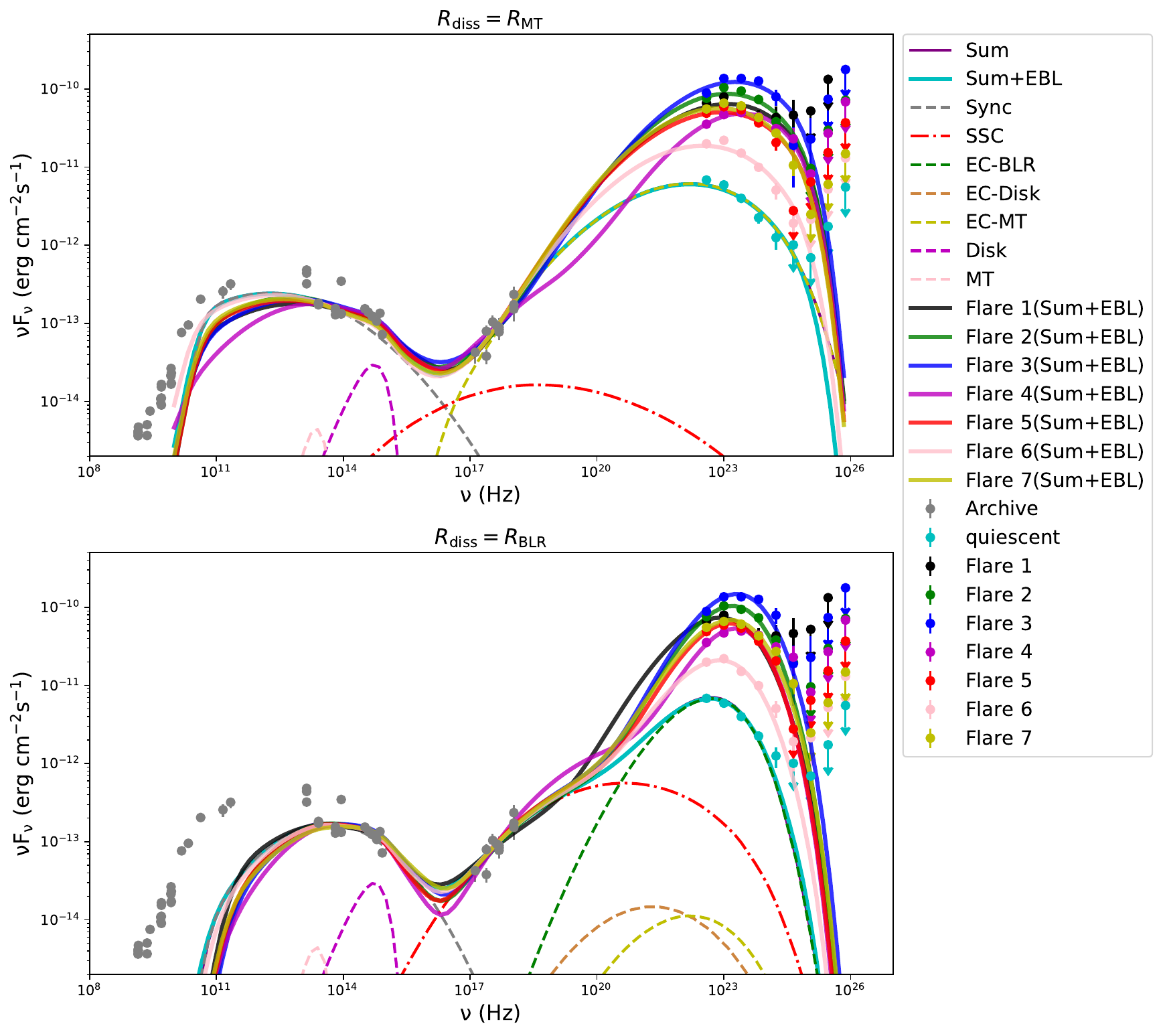}
\caption{Theoretical modeling of the SEDs of B3 1343+451 during the different periods defined in Table~\ref{tab1}. The upper and lower panels represent the best fits with $R_{\rm diss}=R_{\rm MT}$ and $R_{\rm diss}=R_{\rm BLR}$, respectively. The grey points from the radio to X-ray bands represent the historical data obtained from the SSDC Sky Explorer.
Separate synchrotron, MT, accretion disk, SSC, EC-disk, EC-BLR, and EC-MT components are shown. The thick solid line in all plots represents the sum of all components, which have been corrected for EBL absorption considering the model of \citet{Finke2010ApJ}.
\label{fig4}}
\end{figure*}

\subsection{SED Modeling Results\label{sec4.2}}
The model has seven parameters: $B'$, $R'$, $\delta_{\rm D}$, $N_{\rm pk}'$, $\gamma_{\rm pk}'$, $r$, and $R_{\rm diss}$. Note that the absence of the (quasi-)simultaneous data points covering the optical/UV to X-ray frequencies prevent us from effectively estimating the main physical properties characterizing the jet (i.e., $B'$, $R'$, $\delta_{\rm D}$, and $N_{\rm e}'$) through modeling just the SEDs. For FSRQs, the $\gamma$-ray emission is generally attributed to IC scattering off external photons, depending on the distance from the central BH. In our modeling, our main goal is to explore the location of the $\gamma$-ray emission region in the jet, which is crucial for us to understand energy dissipation, production efficiency, and formation mechanisms in relativistic jets.

The large Compton dominance favors two specific locations \citep{Sikora2009ApJ}: (i) inside the BLR or (ii) outside the BLR, but within the MT. The radiation fields of both the BLR and MT remain uniform as long as the emission region is inside the respective component. The precise estimation of the location of the emitting plasma is not very important. To reduce the number of free parameters further, in our analysis we assume $R_{\rm diss}=R_{\rm BLR}$ and $R_{\rm diss}=R_{\rm MT}$. At distances of $R_{\rm diss}=R_{\rm BLR}$, the external radiation is dominated by broad emission lines, while for $R_{\rm diss}=R_{\rm MT}$ the dominant photon field is the IR emission from the MT. Therefore, there are six free parameters in our SED fits. The theoretical SED is calculated using the publicly available JetSet package \footnote{\url{https://jetset.readthedocs.io/en/latest/}} \citep{2009A...501..879T, 2011ApJ...739...66T, 2006A&A...448..861M}. In the SED fits, data above $10^{12}$ Hz are considered, and the model is fitted to this data set using the \texttt{emcee} optimizer\footnote{\url{https://github.com/dfm/emcee}} \citep{2013PASP..125..306F} .

In Fig.~\ref{fig4}, we show the best-fit SEDs with seed photons from the MT (upper panel) and BLR (lower panel). The best-fit values of the parameters and their 1$\sigma$ errors are summarized in Table~\ref{tab3}. It can be seen that the SEDs can be reproduced well by the one-zone leptonic model with a three-parameter LP EED. Moreover, in the BLR fits (i.e., $R_{\rm diss} \sim R_{\rm BLR}$), we note that the X-ray emission is almost entirely attributed to the SSC process, while in the fits with seed photons from the MT (i.e., $R_{\rm diss} \sim R_{\rm MT}$), the X-ray emission is attributed to either the EC mechanism or a combination of EC and SSC mechanisms. The most promising way to distinguish these models is through simultaneous optical to X-ray measurements of the polarization degree and direction \citep[e.g.,][]{Krawczynski2012ApJ,Peirson2018ApJ}. This is mainly because SSC emission can be highly polarized, while in the EC models the unpolarized target photons result in very low polarization degrees.

However, there are no available data from polarization observations for this source. Instead, we determine the type of seed photons by fitting the obtained data to find the best reduced chi-square values relative to the number of degrees of freedom (dof). Based on the $\chi^2_{\rm dof}$  values, we find that the results with seed photons from the MT are significantly better than with seed photons from the BLR. It can also be seen that the EC-MT component better describes the $\gamma$-ray spectral shape than the EC-BLR component. Therefore, our results suggest that the $\gamma$-ray emission may be located outside the BLR, where the external radiation is dominated by IR dust emission.

In MT fittings for different flares, as the peak flux increases, the magnetic field strength and radius show a decreasing trend, while the Doppler factor shows an increasing trend.

Since high-quality continuous multiwavelength observational data are missing for B3 1343+451, we instead focus on the temporal properties of the jet and the origin of the $\gamma$-ray flares to determine the average properties of the jet for this high-redshift blazar. According to our SED modelling, in the MT fits the average values of the magnetic field strength, Doppler factor, and radius are $\langle B'_{\rm MT}\rangle \simeq 0.03$ G, $\langle \delta_{\rm D,MT} \rangle \simeq 37.8$, and $\langle R'_{\rm MT}\rangle \simeq 1.6\times10^{17}$ cm, respectively. In the BLR fits, the values of $\langle B'_{\rm BLR}\rangle$,  $\langle \delta_{\rm D,BLR}\rangle$, and $\langle R'_{\rm BLR}\rangle$ are $\simeq0.7$ G, $\simeq17.4$, and $\simeq9.7\times10^{15}$ cm, respectively. Obviously, the global properties characterizing the jets of the considered source are quite different in the two cases. From Fig.~\ref{fig4}, we can see that the synchrotron bump in the MT fits has a lower peak frequency and more intense radiation in the IR range than found from the BLR fits.

\movetabledown=50mm
\begin{rotatetable}
\begin{deluxetable*}{cccccccccccccccc}
\centering
\tabletypesize{\tiny}
\tablenum{3}
\tablecaption{Parameters Obtained From Modeling the Seven Flares and One Quiescent Period \label{tab3}}
\tablehead{\colhead{$\rm R_{diss}$}&\colhead{Period}& \colhead{$N_{\rm pk}'$}&  \colhead{$B'$}   &\colhead{$\delta_{\rm D}$}&   \colhead{$\gamma_{\rm pk}'$}   &\colhead{r}& \colhead{$R'$} &\colhead{$P_{\rm e}$}&\colhead{$P_{\rm B}$}&\colhead{$P_{\rm r}$ }&\colhead{$P_{\rm p}$ }&\colhead{$P_{\rm jet}$}&\colhead{$\chi^2_{\rm dof}$}\\
\cline{3-14}
\colhead{}&\colhead{}&  \colhead{$\rm \times10^{2} \ [1/cm^{3}]$}    &\colhead{$\rm \times10^{-2}[G]$} & &\colhead{$\rm \times10^{2}$}&   & \colhead{$\rm \times10^{17} \ [cm]$} & \colhead{$\rm \times10^{46} \ [ergs/s]$ } &   \colhead{    $\rm \times 10^{44} \ [ergs/s]$}&\colhead{$\rm \times10^{45} \ [ergs/s]$ }&\colhead{$\rm \times 10^{46} \ [ergs/s]$ }&\colhead{$\rm \times 10^{46} \ [ergs/s]$}
}
\startdata
                &Flare 1  &$2.47^{+0.47}_{-0.46} $& $3.55^{+0.36}_{-0.35} $ & $43.0^{+3.1}_{-2.7} $ & $1.28^{+0.20}_{-0.16} $ & $0.46^{+0.04}_{-0.03}$ & $1.08^{+0.13}_{-0.12}$ & 1.59  & 1.02& 3.51 &7.54& 9.49&1.41\\
                &Flare 2  &$3.32^{+0.57}_{-0.55}$& $3.03^{+0.21}_{-0.20}$ & $44.6^{+2.3}_{-2.1}$ & $1.57^{+0.21}_{-0.18}$ & $0.46^{+0.03}_{-0.02}$ & $0.89^{+0.09}_{-0.09}$ & 1.88  & 0.54& 4.27 &7.39& 9.70&1.50\\
                &Flare 3  &$3.36^{+0.51}_{-0.55}$& $2.38^{+0.20}_{-0.18}$ & $44.1^{+2.0}_{-1.9}$ & $2.05^{+0.35}_{-0.27}$ & $0.44^{+0.03}_{-0.03}$ & $0.85^{+0.10}_{-0.09}$ & 2.12   & 0.30& 5.61 &6.67& 9.35&1.67\\
$R_{\rm diss}=R_{\rm MT}$		&Flare 4  &$1.24^{+0.27}_{-0.23}$& $3.39^{+0.31}_{-0.32}$ & $39.6^{+3.3}_{-3.1}$ & $1.03^{+0.13}_{-0.11}$ & $0.35^{+0.02}_{-0.02}$ & $1.60^{+0.18}_{-0.17}$ & 0.99    & 1.73& 2.85 &7.05& 8.33&1.69\\
                &Flare 5  &$1.78^{+0.29}_{-0.28}$& $3.41^{+0.26}_{-0.23}$ & $37.1^{+2.3}_{-1.8}$ & $1.33^{+0.17}_{-0.15}$ & $0.44^{+0.03}_{-0.02}$ & $1.45^{+0.15}_{-0.13}$ & 1.54   & 1.26& 3.71 &7.29& 9.22&1.49\\
                &Flare 6  &$0.92^{+0.16}_{-0.15}$& $5.15^{+0.38}_{-0.38}$ & $30.4^{+2.1}_{-1.9}$ & $0.92^{+0.12}_{-0.10}$ & $0.46^{+0.02}_{-0.02}$ & $2.46^{+0.22}_{-0.23}$ & 1.13     & 5.56& 2.32 &7.24& 8.66&1.44\\
                &Flare 7  &$2.13^{+0.28}_{-0.25}$& $3.33^{+0.21}_{-0.20}$ & $39.2^{+1.9}_{-2.0}$ & $1.37^{+0.17}_{-0.14}$ & $0.44^{+0.02}_{-0.02}$ & $1.26^{+0.10}_{-0.10}$ & 1.60   & 1.01& 3.77 &7.35& 9.34&1.45\\
                &quiescent&$0.85^{+0.19}_{-0.16}$& $7.47^{+0.58}_{-0.58}$ & $24.1^{+1.7}_{-1.7}$ & $0.62^{+0.09}_{-0.08}$ & $0.44^{+0.03}_{-0.03}$ & $3.52^{+0.40}_{-0.36}$ & 0.94  & 15.04& 1.36 &8.69& 9.92&1.44\\
                & average &$\sim2.01$ &$\sim3.17$ &$\sim37.8$ &$\sim1.27$ &$\sim0.44 $&$\sim1.64$ &$\sim1.47$&$\sim3.31$&$\sim3.43$&$\sim7.40$&$\sim9.25$& \\
\cline{1-14}
\colhead{}&\colhead{}&  \colhead{$\rm \times10^{3}\ [1/cm^{3}]$}    &\colhead{$\rm\times10^{-1} \ [G]$} & &\colhead{$\rm \times10^{2}$}&   & \colhead{$\rm \times10^{15} \ [cm]$} & \colhead{$\rm \times 10^{45} \ [ergs/s]$ } &   \colhead{    $\rm \times10^{43} \ [ergs/s]$}&\colhead{$\rm \times10^{46} \ [ergs/s]$ }&\colhead{$\rm \times10^{44} \ [ergs/s]$ }&\colhead{$\rm \times10^{46} \ [ergs/s]$}\\
\cline{1-14}
                  &Flare 1  &$6.01^{+1.05}_{-0.93}$& $8.36^{+0.76}_{-0.71}$ & $22.7^{+1.2}_{-1.0}$ & $4.01^{+0.56}_{-0.53}$ & $1.01^{+0.11}_{-0.10}$ & $4.17^{+0.53}_{-0.49}$ & 0.93 & 2.34 & 0.87 & 7.62 & 1.04&1.96   \\
                  &Flare 2  &$7.34^{+1.15}_{-0.98}$& $6.70^{+0.55}_{-0.47}$ & $24.4^{+1.1}_{-1.0}$ & $5.21^{+0.50}_{-0.46}$ & $1.03^{+0.09}_{-0.09}$ & $3.34^{+0.34}_{-0.30}$ & 1.11 & 1.12 & 1.00 & 6.90 & 1.18&1.96  \\
                  &Flare 3  &$7.87^{+1.31}_{-1.11}$& $5.72^{+0.43}_{-0.42}$ & $26.4^{+1.2}_{-1.1}$ & $6.51^{+0.66}_{-0.59}$ & $1.18^{+0.12}_{-0.10}$ & $2.77^{+0.30}_{-0.29}$ & 1.29 & 0.66 & 1.14 & 5.95 & 1.33&1.90  \\
$R_{\rm diss}=R_{\rm BLR}$		 &Flare 4  &$2.75^{+0.41}_{-0.38}$& $5.13^{+0.45}_{-0.42}$ & $13.8^{+0.8}_{-0.8}$ & $11.2^{+1.36}_{-1.22}$ & $1.23^{+0.12}_{-0.13}$ & $6.88^{+0.75}_{-0.74}$ & 1.33 & 0.89 & 1.57 & 3.50 & 1.73&2.31 \\
                  &Flare 5  &$3.19^{+0.46}_{-0.43}$& $6.51^{+0.49}_{-0.46}$ & $17.2^{+0.8}_{-0.8}$ & $6.24^{+0.64}_{-0.58}$ & $1.01^{+0.09}_{-0.08}$ & $6.40^{+0.69}_{-0.67}$ & 1.04 & 1.92 & 1.21 & 5.46 & 1.37&1.95   \\
                  &Flare 6  &$1.88^{+0.26}_{-0.24}$& $8.57^{+0.61}_{-0.58}$ & $12.0^{+0.5}_{-0.5}$ & $5.68^{+0.60}_{-0.56}$ & $0.95^{+0.08}_{-0.08}$ & $11.0^{+1.13}_{-1.19}$ & 0.77 & 4.78 & 0.93 & 4.62 & 1.06&2.02  \\
                  &Flare 7  &$4.01^{+0.49}_{-0.45}$& $5.24^{+0.33}_{-0.32}$ & $16.2^{+0.6}_{-0.6}$ & $5.64^{+0.46}_{-0.42}$ & $0.73^{+0.05}_{-0.05}$ & $7.90^{+0.63}_{-0.60}$ & 1.27 & 1.67 & 1.60 & 9.10 & 1.81&2.47  \\
                  &quiescent&$0.43^{+0.07}_{-0.06}$& $8.79^{+0.62}_{-0.60}$ & $6.2^{+0.3}_{-0.3}$ & $7.68^{+1.07}_{-0.95}$ & $0.94^{+0.09}_{-0.08}$ & $34.7^{+3.96}_{-3.75}$ & 0.63 & 13.31 & 1.24& 2.81 & 1.34&1.97  \\
          & average &$\sim4.19$ &$\sim6.88$ &$\sim17.4$ &$\sim6.52$ & $\sim1.01$&$\sim9.65$ &$\sim1.05$&$\sim3.34$&$\sim1.20$&$\sim5.75$&$\sim1.36$& \\
\enddata
\tablecomments{Column 3 denotes the electron density. Column 4 denotes the magnetic field. Column 5 denotes the Doppler factor. Column 6 denotes the peak Lorentz factor. Column 7 denotes the spectral curvature of the EED. Column 8 denotes the radius of the emitting region. Columns 9--12 are, respectively, the power of the jet in the form of relativistic electrons ($P_{\rm e}$), magnetic field ($P_{\rm B}$), radiation field ($P_{\rm r}$), and cold protons ($P_{\rm P}$) with $n_{\rm e}/n_{\rm p}=1$, where $P_{\rm r}$ is the sum of all radiative components. Column 13 is the total power ($P_{\rm jet}=P_{\rm e}+P_{\rm B}+P_{\rm r}+P_{\rm p}$). Column 14 denotes the reduced $\chi^2_{\rm dof} =\rm \chi^2/dof$ derived rom the fits.}
\end{deluxetable*}
\end{rotatetable}

\clearpage

\begin{figure*}[t]
\centering
\includegraphics[width=1\textwidth]{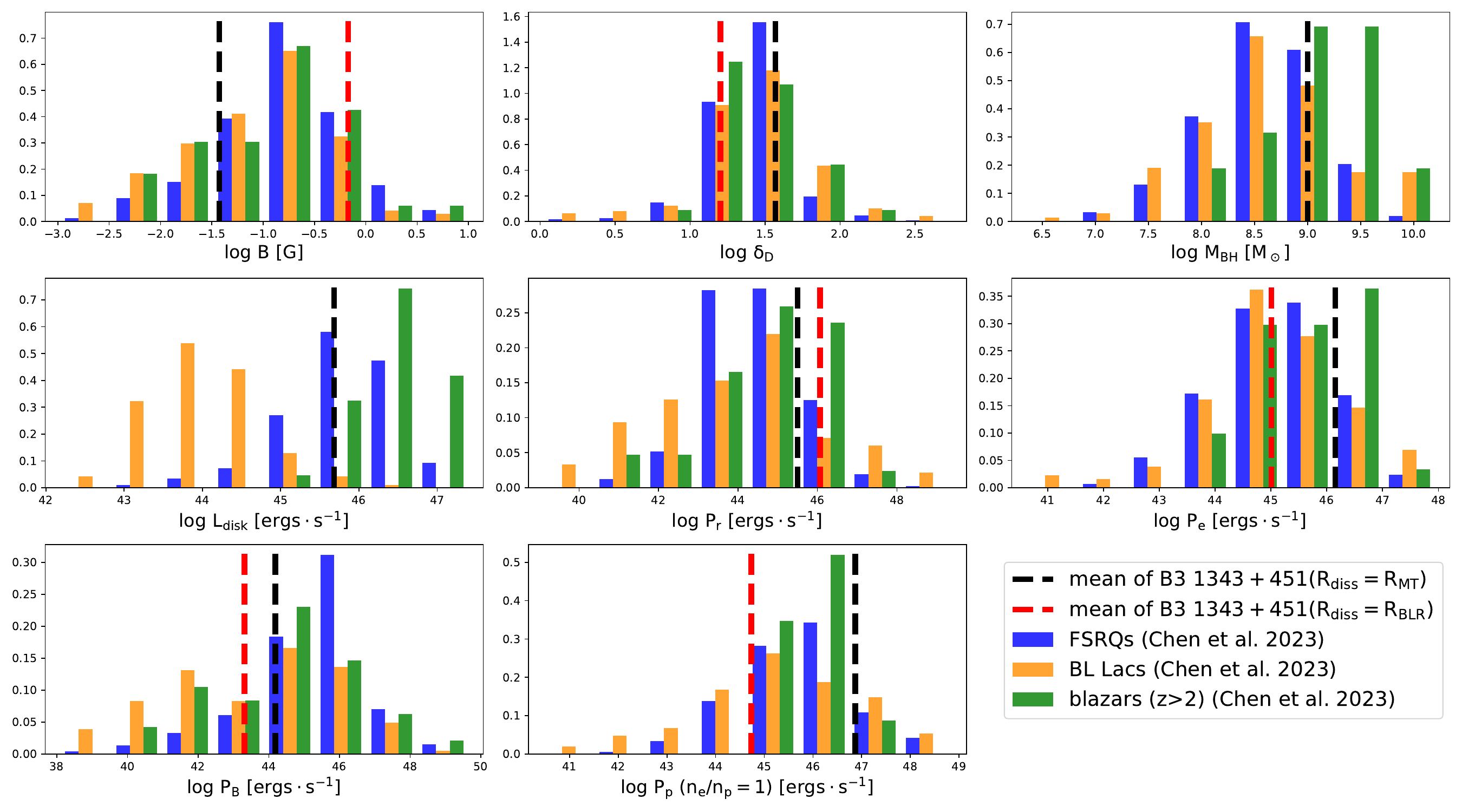}
\caption{The normalized histogram combines the data of 459 blazars from \citet{2023ApJS..268....6C} with our {\bf mean of} parameters, including the magnetic field, Doppler factor, BH mass, and accretion disk luminosity, as well as the radiation, electron, magnetic field, and proton power.
\label{fig5}}
\end{figure*}

\section{Discussion and Conclusion\label{sec5}}
Since the bolometric luminosity of the source is dominated by HE $\gamma$-ray emission, and the $\gamma$-ray luminosity is a tracer of jet power \citep{Nemmen2012Sci,Sbarrato2014MNRAS}, the SED modeling allows us to provide valuable constraints on the power of the jet. Following \cite{Celotti2008MNRAS}, we can evaluate the power carried by the magnetic field ($P_{\rm B}$), emitting electrons/positrons ($P_{\rm e}$), radiation ($P_{\rm r}$), and cold protons ($P_{\rm p}$) in the jet. The derived values for the various epochs can be found in Table~\ref{tab3}. On average, the jet power in the form of emitting electrons/positrons, magnetic field, and radiation field are $\langle P_{\rm e,BLR}\rangle \simeq 1.0\times10^{45}$, $\langle P_{\rm B,BLR}\rangle \simeq 3.3\times10^{43}$, and $\langle P_{\rm r,BLR}\rangle \simeq 1.2\times10^{46}\ \rm ergs\cdot s^{-1}$ in the BLR fits, respectively. In the MT fits, we find $\langle P_{\rm e,MT}\rangle \simeq 1.5\times10^{46}$, $\langle P_{\rm B,MT}\rangle \simeq 3.3\times10^{44}$, and $\langle P_{\rm r,MT}\rangle \simeq 3.4\times10^{45} \ \rm ergs\cdot s^{-1}$, respectively.

In the BLR case, it seems evident that the jet radiative power $\langle P_{\rm r,BLR} \rangle$ cannot be solely ascribed to the overall motion of relativistic electrons/positrons or Poynting flux. Simultaneously, to uphold a globally charge-neutral environment within the blob, the incorporation of a significant proton component is imperative. By considering the protons to be cold and assuming one cold proton per emitting electron, we further obtain $\langle P_{\rm p,BLR}\rangle \simeq 5.7\times10^{44} \ \rm ergs\cdot s^{-1}$, and therefore the total jet power is $\langle P_{\rm jet,BLR}\rangle \simeq 1.4\times10^{46} \ \rm ergs\cdot s^{-1}$. Indeed, the power that the jet expends for producing radiation could be considered to be a lower limit on the total jet power. However, our results indicate that $\langle P_{\rm jet,BLR}\rangle \sim \langle P_{\rm r,BLR}\rangle$ in the BLR case, even if assuming $n_{\rm e}/n_{\rm p}=1$, i.e., a pure $e^-$--$p$ plasma (zero pair content \footnote{The number of $e^\pm$ pairs cannot greatly exceed the amount of protons in order to avoid the Compton rocket effect that would otherwise stop the jet \citep{Sikora2000ApJ,2010MNRAS.405..387G}}), the existence of $e^\pm$ pairs in the jet still reduces the jet power by a factor of $n_{\rm e}/n_{\rm p}$.

This implies that the entire kinetic energy of the jet, defined as $\langle P_{\rm kin}\rangle =\langle P_{\rm e}\rangle +\langle P_{\rm B}\rangle +\langle P_{\rm p}\rangle$, could be used to produce the jet radiation. In this context, less-massive jets (i.e. pair dominated) can be decelerated more effectively than more-massive jets (i.e. with an important proton component), radiation is difficult to sustain . Thus, the HE $\gamma$-ray emission may be predominantly produced by the EC-MT mechanism.

In the MT fits, the kinetic power of jet with the pure $e^\pm$ pairs, $\langle P_{\rm e,MT} \rangle+\langle P_{\rm B,MT} \rangle$, is significantly larger than $\langle P_{\rm r,MT} \rangle$. Assuming an electron-to-proton comoving number density ratio of $n_{\rm e}/n_{\rm p}\sim10-30$ \citep{Kataoka2008ApJ,2010MNRAS.405..387G,Sikora2016Galax,Fan2018ApJ}, we obtain that $\langle P_{\rm p,MT}\rangle$ is in the range of $\sim (2.5-7.4)\times10^{45}\ \rm ergs \cdot s^{-1}$ and therefore $\langle P_{\rm kin,MT}\rangle$ in $\sim(1.8-2.4)\times10^{46} \ \rm ergs \cdot s^{-1}$.

Alternatively, the jet kinetic power can be evaluated through the empirical relation proposed by \cite{Godfrey2013ApJ}:

\begin{equation}
L_{\rm kin,rl}=3\times10^{44}\left(\frac{L_{151}}{10^{32} \ \rm ergs / s \cdot Hz \cdot sr }\right)^{0.67}
\end{equation}

where $L_{151}$ is the extended radio luminosity at 151 MHz. Using the 150 MHz radio flux $F_{150} = 0.23\pm 0.05$ Jy taken from \cite{Shimwell2017}, we find $L_{\rm kin,rl}\simeq6.4\times10^{45}\ \rm ergs\cdot s^{-1}$, which is a factor of $\sim3-4$ lower than the jet kinetic power derived when assuming $n_{\rm e}/n_{\rm p}\sim10-30$.

The discrepancy between these two methods implies that a fraction of the jet energy may be lost after leaving the dissipation region of the blazar. This could be attributed to the formation of the jet’s large-scale structure, which may consume a fraction of the jet energy. As a result, the jet kinetic power, estimated from the extended radio luminosity, is approximately three to four times lower than the value obtained from SED modeling.

Moreover, this discrepancy is further substantiated by numerical modeling of the evolution of radio galaxy lobes, as presented in \cite{Hardcastle2013MNRAS}, where the authors showed that the energy losses of an expanding structure can result in an underestimation of the jet power by about a factor of a few. Motivated by these results, we conclude that the jet power estimated from our EC-MT mechanism may be reasonable.

For a dissipation region located near the MT, the various components powered by the jet of the high-redshift blazar B3 1343+451 can be inferred. As shown in Fig.~\ref{fig5}, we have compared these values with those obtained for the sample of 459 blazars in \citet{2023ApJS..268....6C}. We find that the strengths of these components are significantly more powerful than the obtained averages for the large sample of FSRQs (except for the magnetic field and magnetic field power, which may be attributed to the large errors resulting from the multiwavelength SED fit).
Given the frequent flaring of B3 1343+451, this result is appropriate.
Additionally, we compared our results to a subset of 21 high-redshift (z$>$2) blazars out of 459 blazars. All parameters of the source are fall within the range of high-redshift samples, indicating that B3 1343+451 is a typical high-redshift blazar.

On the one hand, considering that $\langle P_{\rm kin}\rangle$ is greater than $L_{\rm disk}$, it may indicate that the accretion is not enough to explain the jet kinetic power for blazars \citep{2021ApJ...913...93C}. It suggests that the energy may not solely be released into the jet through the Blandford--Payne (BP) process \citep{1982MNRAS.199..883B}. The Blandford--Znajek (BZ) process \citep{1977MNRAS.179..433B} may be nonnegligible.

On the other hand, we find that the total power, including contributions from both a jet and a counterjet, $2\langle P_{\rm jet,MT}\rangle\sim(4.3-5.3)\times10^{46} \ \rm ergs\cdot s^{-1}$, is well below the Eddington luminosity of $ {L}_{\rm Edd}$$\sim1.3 \times10^{47}\ \rm ergs\cdot s^{-1}$ for a BH with a mass of $10^9 M_\odot$, where $2\langle P_{\rm r,MT} \rangle$ is of the same order as $ L_{\rm disk}\sim 4.8\times10^{45} \ \rm ergs\cdot s^{-1}$. These is in agreement with $\langle P_{\rm jet,MT}\rangle$ found by \cite{2014Natur.515..376G} for a large sample of blazars.

It is more interesting to discuss possible accretion structures, we introduce the jet production efficiency is $ \eta=\frac{2\langle P_{\rm jet,MT}\rangle}{\dot{M}c^2}=\frac{2\langle P_{\rm jet,MT}\rangle}{L_{\rm disk}} \eta_{acc}$ (where $ \dot{M}c^2=L_{\rm disk}/\eta_{\rm acc}$, with $ \dot{M}$ as the mass accretion rate and $\bf \eta_{acc}$ as the accretion efficiency),
and \citet{2020MNRAS.495..981S} found $\bf \eta$ for FSRQs $ \sim 0.003 - 0.486$.
To ensure that $\eta$ remains within an appropriate range, we got $\eta_{\rm acc} \sim 0.03\%-5.42\%$. The accretion structure could be a thin disk ($ \eta_{\rm acc}=3\%-30\%$ \citep{2011ApJ...728...98D}, or $\eta_{\rm acc}=5\%-40\%$ \citep{1973blho.conf..343N}); a slim disk ($ \eta_{acc}$ is lower \citep{2005ApJ...628..368O,2007ApJ...660..541G}); or an Advection Dominated Accretion Flow (ADAF) ($\eta_{acc}$ is $< 0.1 \%$ \citep{2009PASJ...61.1313G,2014ARA&A..52..529Y}).

Moreover, $L_{\rm disk}$ is smaller than $L_{\rm Edd}$, indicating sub-Eddington accretion disk \citep{2014ARA&A..52..529Y}, which does not conform to a slim disk. The observations of High Ionization Lines \citep[HILs;][]{1999ApJ...515L..69N}, such as CIV and Ly$\alpha$ by \citet{2020ApJS..249....3A}, are expected to be ionized by a radiatively efficient accretion flow, do not fit with an ADAF. In thin disks, HILs are primarily associated with the high-ionization regions \citep{2000ARA&A..38..521S}, such as accretion disk winds \citep{1995ApJ...448L..81W,2000ARA&A..38..521S}, clouds ionized by the radiation from the accretion disk \citep{2002ApJ...570L..13C}, or the corona above the accretion disk \citep{2017A&A...602L...6V}. So it is more likely that the accretion structure is a standard thin disk.

In conclusion, we propose that the HE $\gamma$-ray emission is likely primarily generated through the EC-MT mechanism, suggesting that the radiative region of B3 1343+451 is situated in close proximity to the MT. We find B3 1343+451 is a typical high-redshift blazar. Except for the magnetic field power, all the other power components are significantly higher than the averages of FSRQs. 
The accretion structure is more likely a thin disk, and jet energy might not only come from the BP process.

\begin{acknowledgments}
We thank the anonymous referee for constructive comments
and suggestions. We acknowledge the  helpful discussion with Prof.
Weimin Gu from Xiamen University. This work is partly supported by the National Science Foundation of China (12263007, 12233006 and 12263003) and the High-level talent support program of Yunnan
Province. We also acknowledge the science research grants from the China Manned Space Project with NO. CMS-CSST-2021-A06.

\end{acknowledgments}

\bibliography{sample631}{}
\bibliographystyle{aasjournal}
\end{document}